\documentclass[prl,twocolumn,showpacs,preprintnumbers,amsmath,amssymb]{revtex4}

\newcommand{\scf}{ScF${_{3} }$}

\usepackage{graphicx}
\usepackage{dcolumn}
\usepackage{bm}

\begin{document}


\title{Large isotropic negative thermal expansion above a structural quantum phase transition}

\author{Sahan U. Handunkanda$^{1}$}
\author{Erin B. Curry$^{1}$}
\author{Vladimir Voronov$^2$}
\author{Ayman H. Said$^3$}
\author{Gian G. Guzm\'{a}n-Verri$^{4,5}$}
\author{Richard T. Brierley$^6$}
\author{Peter B. Littlewood$^{7,8}$}
\author{Jason N. Hancock$^{1}$}

\affiliation{$^{1}$ Department of Physics and Institute for Materials Science, University of Connecticut, Storrs, Connecticut, 06269 USA}
\affiliation{$^{2}$ Kirensky Institute of Physics, Krasnoyarsk, 660036 Russia}
\affiliation{$^{3}$ Advanced Photon Source, Argonne National Laboratory, USA}
\affiliation{$^{4}$ Centro de Investigaciones en Ciencia e Ingenier\'{i}a de Materiales y Escuela de F\'{i}sica, Universidad de Costa Rica, San Jos\'{e}, 2060, Costa Rica}
\affiliation{$^{5}$ Materials Science Division, Argonne National Laboratory, Argonne, Illinois 60349, USA}
\affiliation{$^{6}$ Department of Physics, Yale University, New Haven, CT 06520, USA}
\affiliation{$^{7}$ James Franck Institute, University of Chicago, Illinois 60637, USA}
\affiliation{$^{8}$ Argonne National Laboratory, Argonne, Illinois 60349, USA}

\date{\today}

\begin{abstract}
Perovskite structured materials contain myriad tunable ordered phases of electronic and magnetic origin with proven technological importance and strong promise for a variety of energy solutions. An always-contributing influence beneath these cooperative and competing interactions is the lattice, whose physics may be obscured in complex perovskites by the many coupled degrees of freedom which makes these systems interesting. Here we report signatures of an approach to a quantum phase transition very near the ground state of the nonmagnetic, ionic insulating, simple cubic perovskite material ScF$_3$ and show that its physical properties are strongly effected as much as 100 K above the putative transition. Spatial and temporal correlations in the high-symmetry cubic phase determined using energy- and momentum-resolved inelastic X-ray scattering as well as X-ray diffraction reveal that soft mode, central peak and thermal expansion phenomena are all strongly influenced by the transition.

\end{abstract}
\maketitle

The class of materials with the perovskite structure and chemical formula ABX$_3$ contains examples of perhaps every possible type of physical behavior\cite{Lines1979,Maekawa2004}, much of which is difficult to understand because of the shear complexity of matter. A rich terrain of structural transitions associated with BX$_6$ octahedral tilting in perovskites strongly effects electronic conduction and magnetic exchange pathways, defining the framework of interactions governing a range of physical properties. The A site tolerance appears to be an important parameter in determining the structural phase stability\cite{Lines1979,Maekawa2004,Benedek2013}, but  stable A-site-free perovskites structures are also thermodynamically stable. These are rare cases among oxides (X=O) because the B ions must take on rare hexavalent (+6) electronic configurations, and the only known instance is ReO$_3$. In perovskites based on fluorine (X=F), however, the B ions assume the common trivalent (+3) configuration in an expanded suite of A-site-free perovskite lattices. 

\begin{figure*}
\includegraphics[width=7.in]{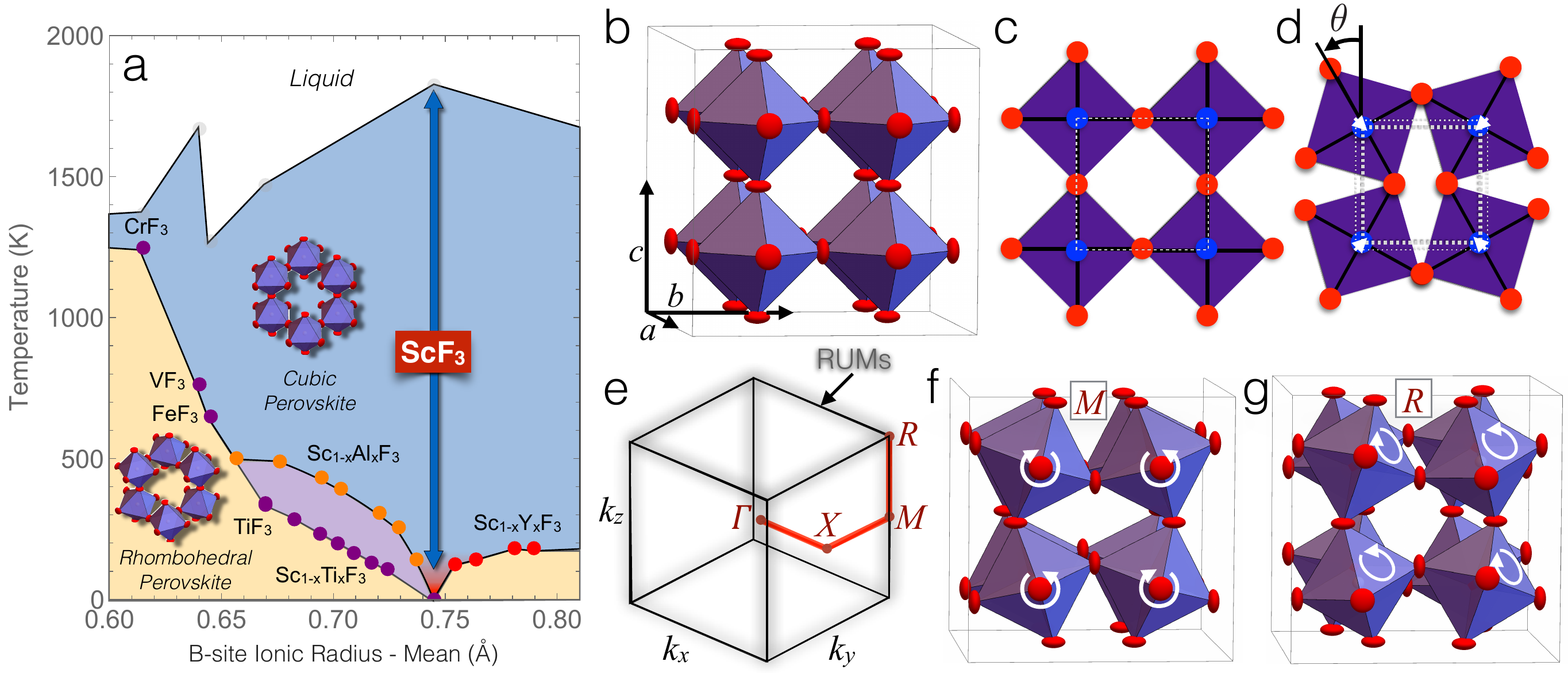}
\caption{
Background material: (a) Structural phase diagram showing the c-r phase boundary versus the B-site mean radius in 3$d$ transition metal trifluroides and in the solid solutions Sc$_{1-x}$Al$_x$F$_3$ and Sc$_{1-x}$Y$_x$F$_3$. Data taken from\cite{Mogus1985,Daniel1990,Morelock2014,Morelock2013b,Morelock2015}. (b) Simple cubic structure of ReO$_3$, ScF$_3$ and the high temperature phases of transition metal trifluorides. Sc$^{+3}$ ions sit at the center of regular corner-linked octahedral cages formed by F$^{-}$ ions (red ellipsoids). (c,d) 100 planar section of the octahedra, illustrating the displacement pattern of modes attributed to NTE. The area of the box in (d) is reduced by $\cos^2\theta$ of the area in (c). (e) The first Brillouin zone of the simple cubic structure, showing the zone center $\Gamma$ (0,0,0), zone face center $X$ ($\pi$,0,0), zone edge center $M$ ($\pi$,$\pi$,0), and zone corner $R$ ($\pi$,$\pi$,$\pi$) points. Soft collective rigid unit modes (RUMs) which nearly preserve internal octahedral bonds are permitted on the zone boundaries as indicated by shadowing. The lowest vibrational modes at the (f) $M$ and (g) $R$ points involve coordinated staggered rotations of octahedra in the patterns shown. The rhombohedral phase can be described as a static tilt according to the $R_4^+$ pattern.}
\label{fig:intro}
\end{figure*}

\begin{figure*}
\includegraphics[width=7.in]{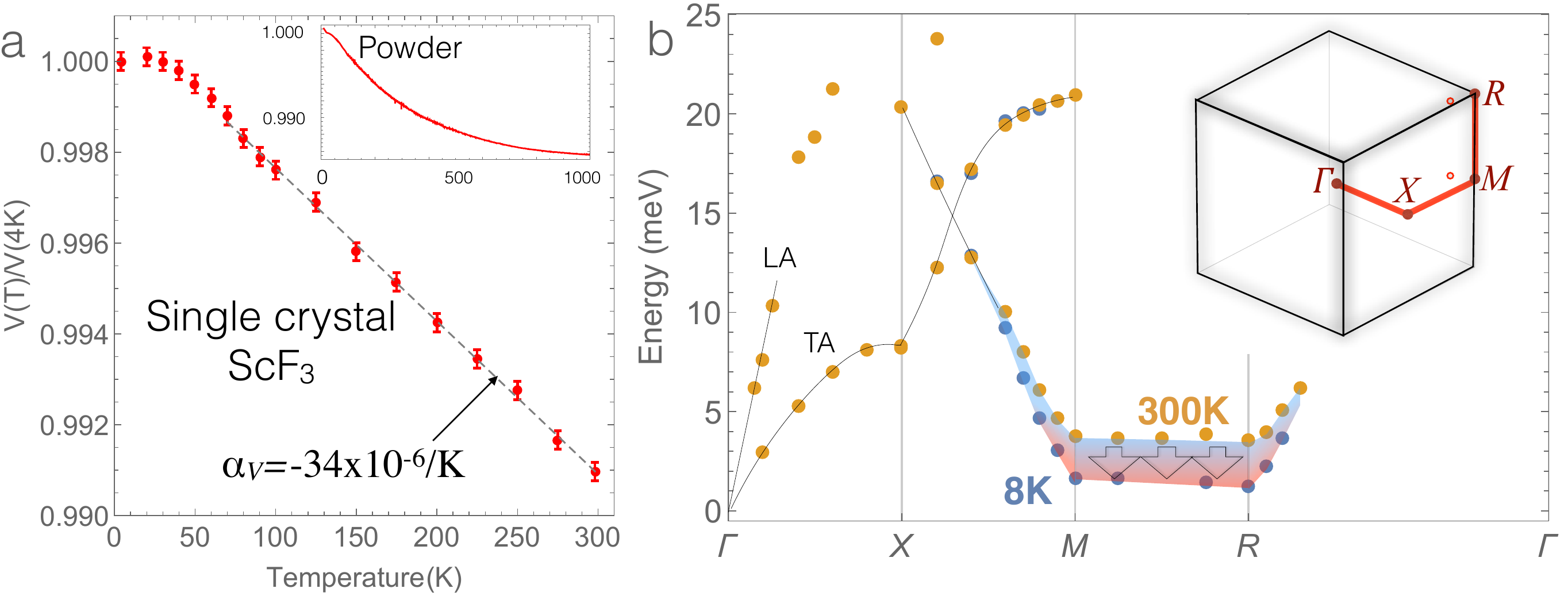}
\caption{
(a) Lattice volume as a function of temperature determined from tracking the Bragg peak (3,0,0) in single crystalline ScF$_3$. Inset: The lattice volume over a larger thermal window determined on powder samples\cite{Greve2010}. (b) Overview of the lattice mode dispersion at select momenta and energies. Strong branch softening along the $M$-$R$ cut is shown by arrows. The longitudinal (LA) and transverse (TA) acoustic branches are shown near $\Gamma$. Black solid lines are a guide to the eye. Red circles in the inset mark the momenta $R+\delta$ and $M+\epsilon$.
}
\label{fig:aTdisp}
\end{figure*}

Figure \ref{fig:intro}a shows a structural phase diagram of BF$_3$ perovskites, where B is a trivalent metal ion\footnote{MnF$_3$ forms a triclinic structure\cite{Hepworth1957}, likely resultant from unusually strong magnetic and Jahn-Teller effects\cite{Hunter2004}. We have not included this in the phase diagram.}. The 3$d$ metal trifluorides display a reversible \cite{Mogus1985} structural cubic-to-rhombohedral (c-r) phase boundary. This sequence of 3$d$ transition metal trifluoride compounds are rhombohedral at room temperature, with the exception of B=Sc, which appears at the zero-temperature terminated c-r phase boundary. Indeed, no rhombohedral phase transition has been observed for \scf\ down to 0.4K \cite{Romao2015}, suggesting that near this composition, the structural phase can be driven by a parameter other than temperature, implying that the ground state of this ionic insulator is very near a quantum phase transition (QPT). Cubic \scf\ further stands out among substances in that it has the most stable structural phase of any known solid trifluoride, retaining high cubic symmetry and a four atom unit cell up to its high melting point $>$1800K\cite{Greve2010,Romao2015}. Separate from the QPT reported here, further interest in this system is due to the recent discovery of extremely large volume negative thermal expansion (NTE) ($\alpha_V\simeq-34\times 10^{-6}/K$ near 300 $K$, Fig. \ref{fig:aTdisp}a) which is both isotropic and thermally persistent from cryogenic (20K) to oven (1100K) temperatures\cite{Greve2010}.

In order to explore this unusual structural quantum material, we have investigated structure and lattice dynamics in high quality single-crystalline \scf\ using X-ray diffraction and inelastic X-ray scattering (IXS). The high momentum resolution (0.02\AA$^{-1}$), nine-analyzer collection scheme, and micro-focused X-ray beam permits measurements of the dynamical structure factor\cite{Said2011} within a transparent, isotopically pure, single-crystalline grain\cite{Aleksandrov2009,Aleksandrov2011} of mosaic width $\simeq$0.002$^\circ$, as measured on the strong $(H,K,L)$=(1,1,0) Bragg reflection indexed to the four-atom simple cubic cell (lattice parameter $a$=4.016$\AA$ at temperature $T$=300K). The dynamical structure factor we have measured is a fundamental property which contains information on the spatial and temporal fluctuations of the lattice degrees of freedom in \scf\ on the approach to the structural quantum phase transition. Figure \ref{fig:aTdisp}b shows an overview of the lowest-energy mode dispersion in \scf\ along high-symmetry directions determined by fitting the structure factor to a damped harmonic oscillator model (Supplemental Information).
Along the cut from the Brillouin zone (BZ) face center $X$ to the BZ edge center $M$, an optic mode crosses the acoustic branches and softens dramatically to a low energy of 3.5 meV at room temperature. Cooling the sample to 8K results in a dramatic and uniform softening of this entire $M$-$R$ branch to $\sim1meV$, further suggesting an approach to a structural instability near zero temperature. 

The low $M$-$R$ branch is a near-degenerate manifold of cooperative vibrational modes which are soft because they largely preserve the internal dimensions of the stiff metal-anion molecular sub-units. In order to retain local constraints of bond distances and angles, staggered rotations induce a shrinking of the cell dimensions (Fig \ref{fig:intro}c,d shows this for the $M_3^+$ distortion), establishing a long-range coupling between local transverse linkage fluctuations and lattice volume that has long been ascribed as the cause of structural NTE\cite{Pantea2006,Grima2006,Lind2012,Takenaka2012a}. 
$M$-$R$ is an important mode branch in perovskites because temperature-driven condensation of modes on this branch can describe many structural transitions which accompany ordering transitions of other types\cite{Benedek2013,Maekawa2004,Lines1979}. These soft modes circumscribe the BZ edges (Figure \ref{fig:intro}e) within a very narrow window of energy set by the $M$-$R$ branch dispersion, with the $R$ point $\simeq$200$\mu$eV lower at all temperatures measured, consistent with an approach to the symmetry-lowered rhombohedral $\overline{R}3c$ structure, as observed in other 3$d$ BF$_3$ systems.

The rhombohedral $\overline{R}3c$ structure is related to the high temperature cubic $Pm\overline{3}m$ structure via a staggered octahedral tilt around the $\langle 111 \rangle$ axis (Figure \ref{fig:intro}a inset), corresponding to a frozen $R_4^+$ lattice distortion (Figure \ref{fig:intro}g). The staggered 111 octahedral rotation angle is therefore the order parameter of the rhombohedral phase. The free energy curvature and origin of cubic-phase stabilization in trifluorides has been addressed in a body of theoretical and computational work aimed to generally understand perovskite structural phases. Density functional theory\cite{Boyer2007}, molecular dynamics\cite{Chaudhuri2004}, and exact electrostatic energy considerations\cite{Allen2006} of the insulating AlF$_3$ prototype trifluoride suggest that Madelung energy including static electric dipole-dipole and induced dipole-induced dipole interactions compete to determine stability of the cubic phase, and results in a low temperature rhombohedral phase. It has been pointed out\cite{Allen2006} that these competing influences could cancel identically for larger metal radii, marginally stabilizing the cubic phase for all temperatures and it appears this limit is realized for ScF$_3$. This suggestion is supported by related previous work using momentum-integrated time-of-flight neutron spectroscopy on bulk powder samples of \scf\ showed temperature dependence of high energy $>$10meV peaks in the neutron-weighted density of states and included calculations suggesting the curvature of the $R_4^+$ mode is quartic in the tilt angle\cite{Li2011}. In what follows we present a momentum, energy, and temperature-resolved investigation of the approach to the quantum phase transition between the simple cubic and rhombohedral phases and its effects on the lattice dynamics of perovskite ScF$_3$.

\begin{figure}
\includegraphics[width=3.4in]{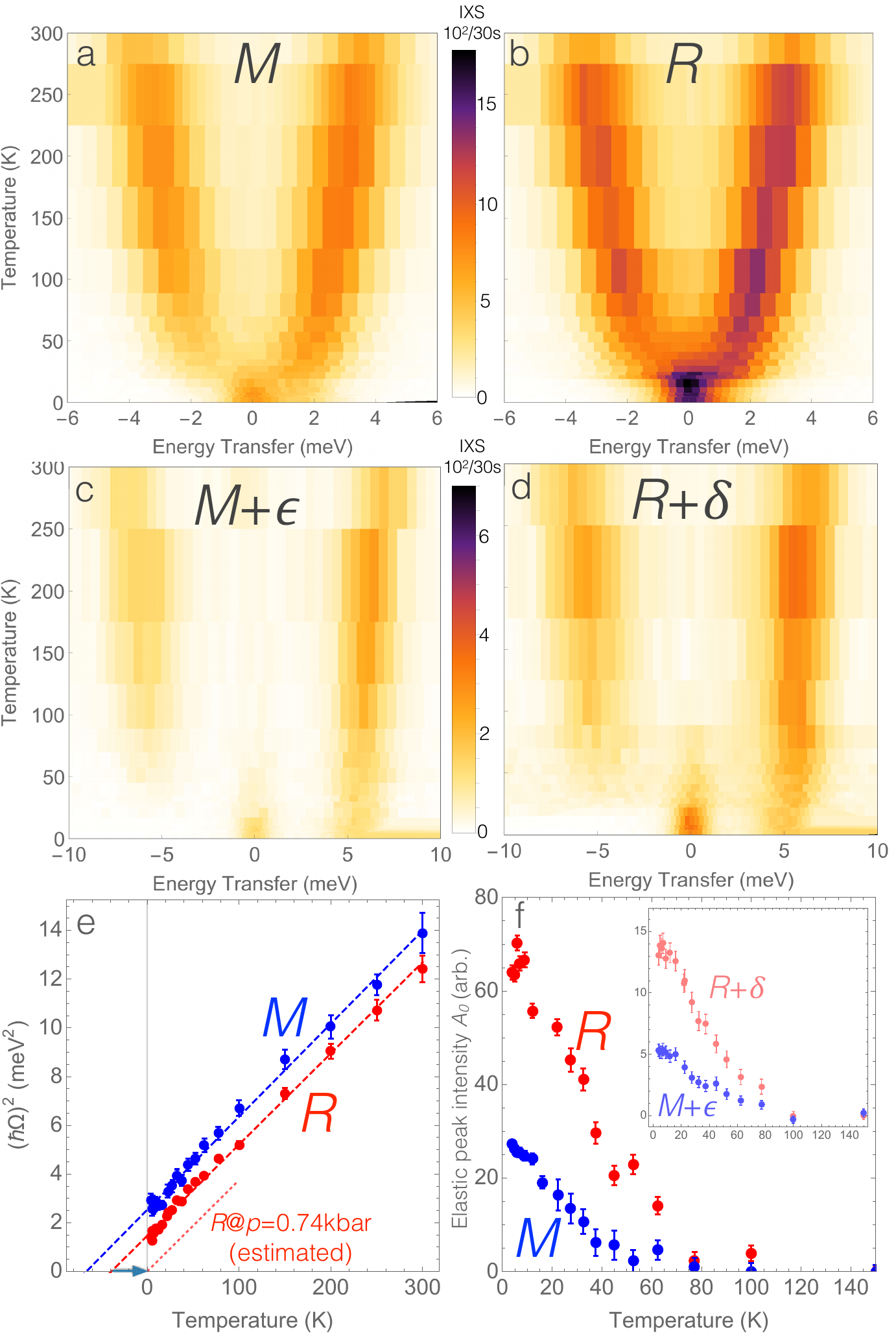}
\caption{
Color maps of the IXS signal at the (a) $M$, (b) $R$, (c) $M$+$\epsilon$ and (d) $R$+$\delta$ points in reciprocal space. Resolution-convoluted fitting of the IXS data in (a)-(d) results in squared mode energies (e) and elastic peak intensity (f). Dashed lines in (e) show the extrapolation of the mode frequencies from classical mean field theory. Dotted lines show the expected effect of a small 74MPa pressure on the $R$ point mode energy, suggesting this small pressure is necessary to pass through the QPT.}
\label{fig:ixs}
\end{figure}

Figures \ref{fig:ixs}a,b show the temperature-dependent spectra at $M$ (2.5,1.5,0) and $R$ (2.5,1.5,0.5) points, where the dynamical structure factor is particularly strong for the modes of Figure \ref{fig:intro}f,g. Clear soft mode behavior is observed concomitant with the emergence of a strong elastic peak, which onsets at $\sim$80K and strengthens dramatically as the temperature is lowered. The elastic peak emergence is also observed at momenta away from the high-symmetry $M$-$R$ cut from data collected in an adjacent analyzer channel. Figure \ref{fig:ixs}c,d show IXS data at $M$+$\epsilon$ (2.39,1.43,-0.02) and $R$+$\delta$ (2.39,1.43,0.45) momentum points, where the optical branch is much stiffer (5-6meV) and displays a milder degree of softening, making the distinction between the elastic peak and soft mode phenomena more apparent in the raw data. In all cases, a transfer of spectral weight from the optical mode to the central peak feature is observed as the temperature is lowered.

All data were fit using a resolution-convoluted model for the dynamical structure factor which includes a damped harmonic oscillator and detailed balance condition to describe the lattice mode, and a resolution-limited elastic `central' peak, consistent with previous analysis of this phenomena in SrTiO$_3$ and other materials (see Supplemental Information)\cite{Shapiro1972}. This analysis allows us to extract the soft mode frequency $\Omega$ and strength $A_0$ of the central peak. Figure \ref{fig:ixs}e shows that the temperature dependence of the lowest soft-mode frequency behaves approximately as $\Omega^2 \propto (T-T_c)$, as expected
from classical mean-field theories of structural phase transitions\cite{Cowley1980}.
This simple extrapolation of the $R$ point soft mode frequency suggests
that the phase transition occurs at $T_c \simeq -39K$,
so does not occur for any finite temperature, consistent with observations\cite{Greve2010,Morelock2014,Morelock2013b,Morelock2015}.

Central peak and mode softening were discovered in the context of the 110K
cubic-to-tetragonal transition of incipient
ferroelectric SrTiO$_3$\cite{Riste1971,Shapiro1972,Muller2010,Kisiel2015a}, but 
are commonly associated with an approach to a structural phase transition.
There, the elastic central peak has maximum strength at $T_c$
but begins to appear up to 25K higher\cite{Shapiro1972,Kisiel2015a}.
In the present case, we observe no maximum in elastic
scattering at any finite temperature and the intensity continues to strengthen at the lowest temperatures measured (3.8 $K$).
In \scf, the effect of the central peak begins above 80K, suggesting that precursor effects occur as high as 120K above our extrapolated transition temperature. The thermally robust central peak in ScF$_3$ occurs in our strain-free, (naturally) isotopically pure, color center-free single crystals with narrow (0.002$^\circ$) mosaic. Interestingly, the elastic peak begins to strengthen at the same temperature scale at which the thermal expansion saturates, discussed further below.

Though central peaks have been observed in nearly all systems exhibiting structural phase transitions,
their origins are still under debate\cite{Cowley2006}.
While one class of theories have proposed that the central peak is the result of intrinsic nonlinear mechanisms
such as the presence of solitons\cite{Krumhansl1975}, another class of theories suggest that it is an extrinsic effect due to defects\cite{Cowley2006}. Based on the appearance of the central peak at a broad manifold of momenta throughout the Brillouin zone, we have observed localized excitations in the fluctuation spectrum below 100K. Below $T_c$ in related material TiF$_3$\cite{Mogus1985}, needle-like 111 oriented domains proliferate throughout the material, and we speculate that the localization of excited states may be a precursor effect related to the formation of the domain structure in the ordered state. 

\begin{figure}
\includegraphics[width=3.in]{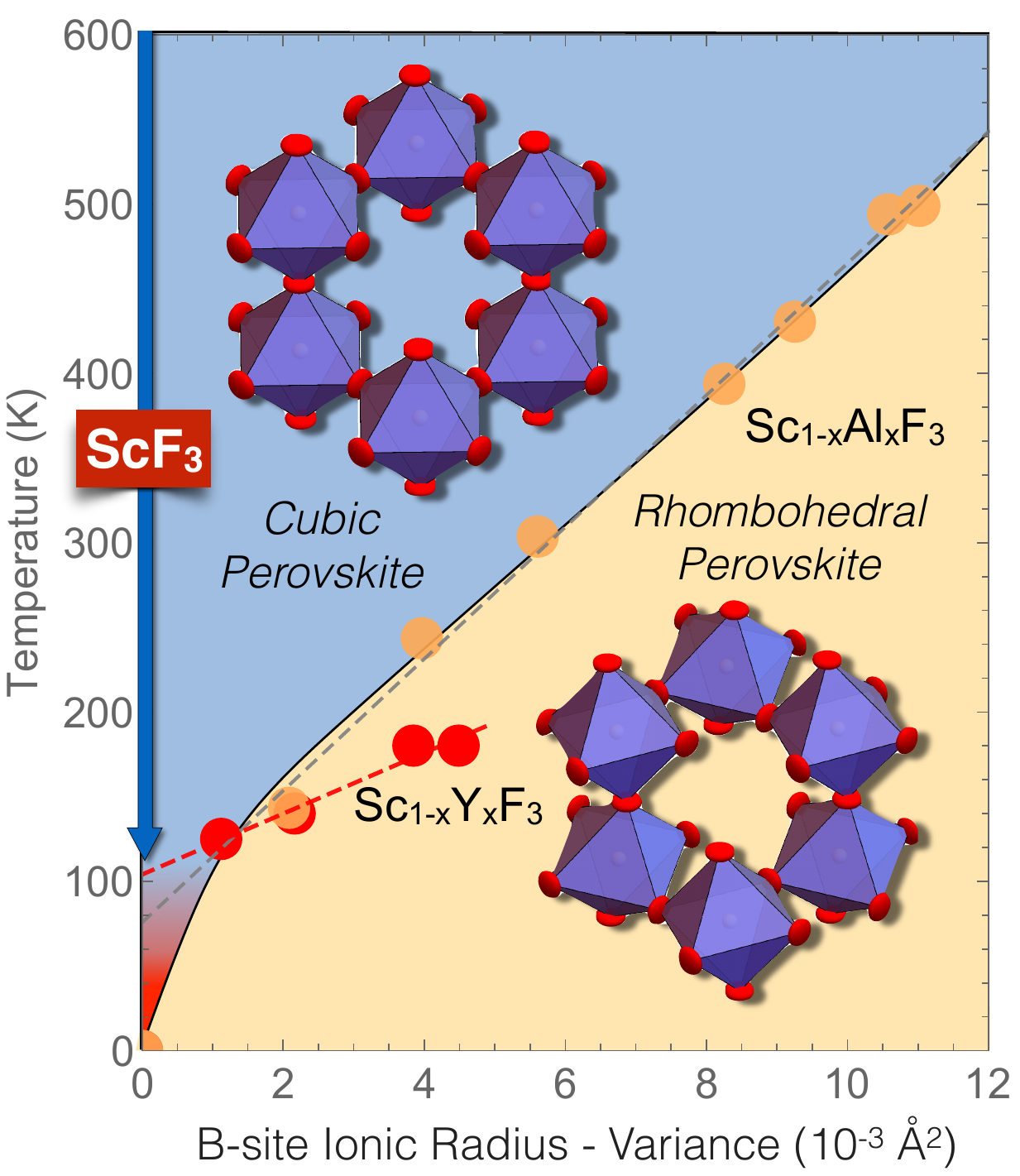}
\caption{
Disorder phase diagram for \scf. Compositional disorder is quantified via the B-site variance $\sigma_B^2=\langle r_B-\langle r_B\rangle\rangle^2$, which has the effect to increase the stability of the lower-symmetry, rhombohedral structural phase. Red and orange symbols show the effect of Y and Al substitution, respectively. Dotted lines indicate a linear fit to the $T_c(\sigma^2_B)$, not including the pure sample \scf. The vertical axis intercept reveals a temperature scale $\simeq$80-100 $K$ where central peak elastic scattering and thermal expansion saturation are also observed in pure \scf. 
}
\label{fig:variance}
\end{figure}

The c-r structural transition can be stabilized at finite temperature through substitution of other metals (Al,Ti,Y) for Sc, or in pure \scf\ through application of very low hydrostatic pressures\cite{Aleksandrov2009,Aleksandrov2011} ($<$0.2GPa) at cryogenic temperatures. Remarkably, using the observed pressure dependence of the transition temperature $dT_c/dp\simeq$525K/GPa\cite{Aleksandrov2009,Aleksandrov2011,Greve2010}, and the putative transition temperature $T_c\simeq$-39K, one can estimate that pressures as small as 740bar=0.074GPa would be sufficient to drive the transition upward to 0K, permitting a clean manner in which to observe
a structural quantum phase transition. The sensitivity of the phase boundary to these external changes suggests that the nature of the cubic phase is delicate at low temperature and susceptible to even mild perturbations. 
Chemical substitution of Sc by Ti\cite{Morelock2014}, Al\cite{Morelock2015}, and Y\cite{Morelock2013b} have been reported and $T_c$ is presented in Fig \ref{fig:intro}a. The phase boundary in Sc$_{1-x}$Ti$_x$F$_3$ appears to linearly interpolate between the end members ScF$_3$ and TiF$_3$ with no maxima or minima at intermediate compositions, consistent with a picture where substitution of well-matched ionic radii ($r_{Ti}$=0.670$\AA$; $r_{Sc}$=0.745$\AA$) uniformly affects the potential landscape and gradually changes the phase stability. 

On the other hand, substitution of ions with radii much smaller $r_{Al}$=0.535$\AA$ or much larger $r_{Y}$=0.90$\AA$ than Sc stabilizes the rhombohedral phase, suggesting the role of disorder is strong for these substitutions. We quantify the disorder in these solid solutions using the variance of the B-site ion distribution $\sigma^2_B$=$\langle  r_B$-$\langle r_B\rangle \rangle^2$ as determined from the nominal chemical formula in these solid solutions and plot the disorder phase diagram in Figure \ref{fig:variance}. The trend in $T_c(\sigma_B^2)$ is an example of the general case, wherein electronic (superconducting, magnetic) and structural (ferroelectric, ferroelastic) transitions have opposite sensitivities to disorder, as determined through studies of the A-site variance effect on phase stability\cite{Rodriguez-Martinez1996,Attfield2001}. The latter case, relevant here, occurs because local strain associated with substitution has the effect of disrupting long range propagation of strain, enhancing the stability of the lowered-symmetry phase. Generalizing these trends to B-site disorder near the composition \scf, we separately fit the transition temperatures $T_c(x)$=$T_c^0$-$p_1 \sigma_B^2$ to the doped compositions of Al and Y substitution and find that extrapolating transition temperatures of the doped compositions to the ideal case $\sigma_B^2$=0 gives roughly similar $T_c^0$ = 75K (Al) and 104K (Y), which is near the temperature scale where we observe strong elastic peak emergence (Fig 3f) and the thermal expansion coefficient departs from linear dependence on temperature (Fig 2a). Empirically $T_c<0.4K$ is observed for pure \scf, suggesting strong deviation from previous studies\cite{Rodriguez-Martinez1996,Attfield2001} and qualitatively different physics in the pure limit. 

The ScF$_6$ octahedra in pure \scf\ move collectively in a state of strong orientational fluctuation\cite{Li2011,Allen2006}, the strength of which can be estimated from powder refinement of the anisotropic thermal parameter $U_{33}$, associated with transverse F motion. The orientational fluctuations are significant\cite{Greve2010}, approaching $2\sqrt{U_{33}}/a\simeq$9.2$^\circ$ FWHM at room temperature, significantly larger than the isostructural oxide ReO$_3$\cite{Rodriguez2009}. In this regime, the influence of long-ranged strain draws in the lattice, leading to the robust and sizable NTE effect. At elevated temperatures, long-range strain fields imposed by local constraints propagate over much shorter distances and the fluctuation spectrum lacks a central peak and appears more conventional. The long-range strain effects we have observed are likely present in the physics of structural transitions in perovskite and other structural classes, but are particularly enhanced in \scf\ at low temperature.

In conclusion, we have observed signatures of an approach to a structural phase transition occurring close to zero temperature and pressure in the strong negative thermal expansion materials ScF$_3$. We have identified the zone-boundary soft mode branch associated to an incipient structural transition in ScF$_3$ together with a thermally robust, resolution-limited central peak. Analysis of the lattice instability suggests that weak pressures are sufficient to induce the c-r transition very near zero temperature. These observations indicate that ScF$_3$ is a candidate material to explore the effects of quantum mechanics on central peaks and structural phases.

Work at the University of Connecticut is supported by National Science Foundation award DMR-1506825.
Work at the University of Costa Rica is supported by
Vicerrector\'{i}a de Investigaci\'{o}n 
under the project no. 816-B5-220, 
and work at Argonne National Laboratory is
supported by the U.S. Department of Energy, Office of
Basic Energy Sciences under contract no. DE-AC02-06CH11357. RTB acknowledges support from the Yale Prize Postdoctoral Fellowship. The construction of HERIX was partially supported by the NSF under Grant No. DMR-0115852.

\bibliography{library}

\end{document}